\documentclass[aps,prl,twocolumn]{revtex4}

\usepackage{tabularx}
\usepackage{amsmath}
\usepackage{amssymb}
\usepackage{graphicx}
\usepackage{dcolumn}
\usepackage{bm}

\begin{document}
\title{Exciton polarization, fine structure splitting and quantum dot
asymmetry under uniaxial stress}
\author{Ming Gong}
\author{Weiwei Zhang}
\author{Guang-Can Guo}
\author{Lixin He}
\email{helx@ustc.edu.cn}
\affiliation{Key Laboratory of Quantum Information, University of Science and Technology of 
China, Hefei, 230026, Anhui, People's Republic of China}

\date{\today}

\begin{abstract}
We derive a general relation between the fine structure splitting (FSS) and the
exciton polarization angle of self-assembled quantum dots (QDs) under uniaxial stress. 
We show that the FSS lower bound under external stress 
can be predicted by the exciton polarization angle and FSS under zero stress.
The critical stress can also be determined by monitoring the change in
exciton polarization angle.
We confirm the theory by performing atomistic pseudopotential calculations 
for the InAs/GaAs QDs.
The work provides a deep insight into the dots asymmetry and their optical
properties, and a useful guide in selecting QDs
with smallest FSS which are crucial in entangled photon sources
applications. 

\end{abstract}

\maketitle

The recent successful demonstration \cite{stevenson06}  of entangled photon 
emission from a biexciton cascade process \cite{benson00} in a single quantum dot
(QD) represents a significant advance in solid state quantum information applications.
The ``on-demand'' QD entangled photon source has fundamental advantages 
over traditional sources using spontaneous 
parametric down conversion process which are probabilistic \cite{gisin02}.
Furthermore, QD emitters can be driven electrically rather than 
optically, and therefore have many advantages in device applications 
\cite{salter10}. However, it is still a big challenge in the generation of 
high-quality entangled photon pairs from QDs.

\begin{figure}
\centering
\includegraphics[width=2.8in]{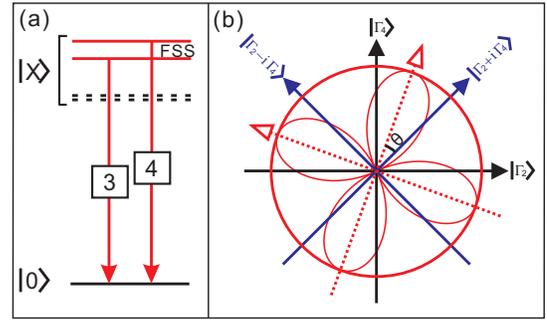}
\caption{(Color online) (a) Four lowest energy levels of monoexciton in
a QD, including two dark states (broken lines) and two bright states (solid lines).
The energy difference between the two bright states defines the FSS. 
(b) Polarization of exciton emissions under different 
symmetry. The black line corresponds to $H_{2v}$ without spin-orbit coupling, the blue 
line corresponds to $H_{2v}$ with spin-orbital coupling, and the red dotted line 
corresponds to $C_1$ symmetry.}
\label{fig-symmetry}
\end{figure}

The key issue here is to suppress the fine structure splitting (FSS) 
[Fig. \ref{fig-symmetry} (a)] 
of the monoexciton, which arises from the underlying asymmetry of the QDs.
There have been many attempts to reduce the FSS in QDs, including thermal
annealing \cite{tartakovskii04, stevenson06}, 
choosing proper dot matrix materials \cite{he08}, and growing the dots 
in high symmetry directions \cite{singh09}. However, these methods can only reduce the FSS to
the level of about 10 $\mu$eV, which might still be too large to produce high-quality
entangled photon pairs. 
It has been shown that the FSS can be effectively tuned by an external magnetic
field \cite{stevenson06} and electric field \cite{gerardot07,bennett10}.
Perhaps a more convenient way to tune the FSS in a QD is via uniaxial
stresses \cite{seidl06}. 
Singh {\it et al} \cite{singh10} showed that the FSS can be tuned to zero under
uniaxial stress for an ideal QDs with $C_{2v}$ symmetry. However, for a general
QD, which has $C_1$ symmetry, there is a lower bound for the FSS when
an external stress is applied. 
It was unclear which or what kind of QDs may have 
the smallest FSS lower bound under stress.  
Therefore, it is still an open question in selecting QDs
that are suitable for entangled photon emitters.

The purpose of the letter is to establish a general relationship between the
asymmetry in QDs, the exciton polarization angle and the FSS under uniaxial stress,
therefore provides a useful guide in selecting QDs that have the smallest
FSS lower bounds for applications such as entangled photon emitters. 
We show that QDs in which the exciton has polarization closely
aligned along the [110] or [100] directions have the smallest FSS under stress.
The critical stress can also be determined by monitoring the change 
in the exciton polarization angle.
The theory is further confirmed
through an atomistic empirical pseudopotential calculations for InAs/GaAs QDs.

We start from a general understanding of the relation between the QD
symmetry and exciton polarization angles shown in Fig. \ref{fig-symmetry}(b). 
An ideal InAs/GaAs QD has $C_{2v}$ symmetry. 
Without spin-orbit coupling, the two bright exciton states belong to two different 
irreducible representations $|\Gamma_2\rangle$ and
$|\Gamma_4\rangle$. 
The polarizations of the two bright states are along the [100]
and [010] directions respectively.
When the spin-orbital interaction is included, the two bright states belong to
irreducible representation 
$|\Gamma_2\rangle \pm i|\Gamma_4\rangle$ \cite{bester03}, therefore the polarizations of the emission
lines should be exactly along the $[110]$ and $[1\bar{1}0]$ directions.
For a general dot, the symmetry is further lowered to $C_1$ owing to
structural imperfections or alloy randomness \cite{mlinar09},
the polarization angle will depart from the 
$[110]$ and $[1\bar{1}0]$ directions, i.e., $\theta \neq$0.

When uniaxial stress is applied, 
the exciton Hamiltonian can be written as,
\begin{equation}
H({\bf n}, p) = H_{2v} + V_{1} + V_s({\bf n})p \, ,
\end{equation}
where ${\bf n}$ is the external stress direction,
and $p$ is the magnitude of the stress. 
$H_{2v}$ represent the Hamiltonian of an idea QDs with $C_{2v}$ symmetry, 
whereas $V_1$ lower the dot symmetry to $C_1$, 
due to local structure deformations, alloy distribution \cite{mlinar09} and interfacial 
effects \cite{bester05a} etc. $V_s({\bf n})p$ is the potential change due to
the external stress.
We neglect the higher-order $O(p^2)$ terms. This is justified by
atomistic pseudopotential calculations, which show that these terms
are negligible up to 200 MPa. 
The eigenvectors of the two bright states of $H_{2v}$ are 
$|3\rangle = |\Gamma_2\rangle + i|\Gamma_4\rangle$ and
$|4\rangle = |\Gamma_2\rangle - i|\Gamma_4\rangle$, with corresponding eigenvalues $E_3$ 
and $E_4$, respectively. 
The energy levels are schematically shown in Fig. \ref{fig-symmetry}(a) in solid
lines. The difference $\Delta = |E_3 - E_4|$ is the FSS. The other two states
$|1\rangle$ and $|2\rangle$ are optically dark,
also shown in Fig. \ref{fig-symmetry}(a) in broken lines.
In the absence of an in-plane magnetic field, the coupling between the dark
states and bright states is negligible.
We therefore write the Hamiltonian in the space spanned by the two bright states,
\begin{equation} 
H = \begin{pmatrix} \bar{E} + \delta +\alpha_3 p & \kappa +\beta p  \\
						  \kappa +\beta p & \bar{E} - \delta  + \alpha_4 p	
\end{pmatrix},
\label{eq-22}
\end{equation}
where $\bar{E}+ \delta = \langle 3|H_{2v}+V_1|3\rangle$, $\bar{E} - \delta =\langle 4|H_{2v}+V_1|4\rangle $.
$\alpha_i = \langle i|V_s({\bf n})|i\rangle$ ($i$=3, 4), $\kappa = \langle 3|V_{1}|4\rangle$ and 
$\beta = \langle 3|V_s({\bf n})|4\rangle$. 
Because the Hamiltonian has a time-reversal symmetry,
all parameters can therefore be set to real values for simplicity. 
Diagonalization of the Hamiltonian yields eigenvalues,
\begin{equation}
E_{\pm} = \bar{E} + {p\gamma \pm \sqrt{4(\beta p+\kappa)^2 + (\alpha p + 2\delta)^2} \over 2},
\label{eq-E}
\end{equation}
where $\alpha = \alpha_3 - \alpha_4$ and $\gamma = \alpha_3
+ \alpha_4$. The eigenvectors of the two states are 
$|\psi_{\pm}\rangle =  -{-2\delta -p\alpha \pm \sqrt{4(\beta p+\kappa)^2 + 
(\alpha p + 2\delta)^2} \over 2(\beta p+\kappa))} |3\rangle + |4\rangle$. 
Since all parameters are real, the two states are linearly polarized 
\cite{footnote1}. 
We calculate $\Delta(p)$ for QDs under uniaxial stress $p$,
\begin{equation}
\Delta (p) = \sqrt{4(\beta p+\kappa)^2 + (\alpha p + 2\delta)^2}\, .
\label{eq-FSS}
\end{equation}
Because $\Delta^2$ is a quadratic function of $p$, 
the lower bound of FSS and the corresponding critical stress can 
be calculated analytically,
\begin{equation}
\Delta_c = {2|\alpha\kappa - 2\beta\delta|\over \sqrt{\alpha^2 + 4\beta^2}},\quad
p_c= -2{\alpha\delta +2\beta\kappa \over \alpha^2 + 4\beta^2}.
\label{eq-critical}
\end{equation}
The polarization angle $\theta$ vs $p$ can be calculated from, 
\begin{equation}
\tan(\theta_{\pm}) = { -2\delta -p\alpha \mp \sqrt{4(\beta p+\kappa)^2 + (\alpha p + 2\delta)^2} 
\over 2(\beta p+\kappa)} \, .
\label{eq-theta}
\end{equation}
Obviously, $\theta$ changes with $p$.
At the critical stress $p_c$, we have $\tan(\theta_c^{\pm}) = 
{2\beta \over \alpha} \mp \sigma \sqrt{1+({2\beta \over \alpha})^2}$,
where $\sigma$ is the sign of $(-2\beta\delta +\alpha\kappa)$. It is interesting to find that
the polarization angle at the critical stress is independent of $\delta$ and $\kappa$, but
only on the ratio of $\beta/\alpha$. Thus the values of $\alpha$, $\beta$, 
$\kappa$ and $\delta$ can be uniquely determined using the relationship
between $\Delta$ and $p$, and the polarization angle at $p_c$.

\begin{figure}
\centering
\includegraphics[width=3.3in]{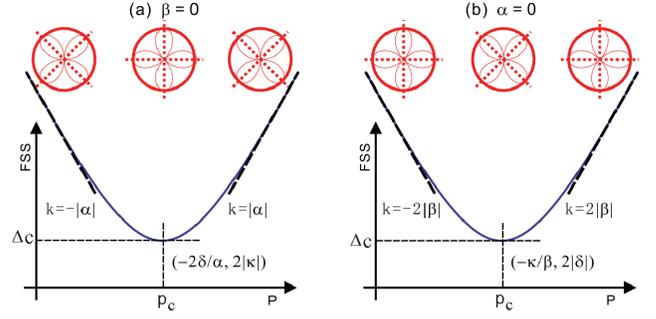}
\caption{(Color online) FSS vs $p$ for (a) $\beta = 0$ and (b) $\alpha = 0$. 
The circles above show the exciton polarization angles at $p_c$ and far away from $p_c$.
}
\label{fig-lowbounds}
\end{figure}

For a dot with $C_{2v}$ symmetry, the external stress along the [110] and [1$\bar{1}$0] directions 
would not change the symmetry of the dot. For a general dot, the symmetry of
the dot is lowered to $C_1$ due to alloy distribution and structural
asymmetry \cite{mlinar09, bester05a}, which change little under external stress. Therefore, for
stress along the [110] and [1$\bar{1}$0] directions, we have $\beta \approx$0. 
Another special case arises when stress is applied along the [100] direction, for which
we have $\alpha$=0 by symmetry. The results for these two special cases are
schematically shown in Fig. \ref{fig-lowbounds} (a), (b) respectively.
In Fig.~\ref{fig-lowbounds}(a), the change in FSS with $p$ is determined
by $|\alpha|$, and the lower bound is determined by $2|\kappa|$ at $p_c =
-2\delta/\alpha$. 
At $p = p_c$, the polarizations are along the [100] or [010] directions whereas far away
from $p_c$, the polarizations are along the [110] or [1$\bar{1}$0] directions.
The results for $\alpha = 0$ shown in 
Fig. \ref{fig-lowbounds} (b) are totally different. 
In this case, the change in FSS with $p$ is determined by $2|\beta|$, and 
the lower bound is $2|\delta|$ at $p_c = -\kappa/\beta$. 
At $p = p_c$, the polarization is along the [110] and [1$\bar{1}$0] directions, 
whereas far away from $p_c$, the polarization is rotated
into the [100] and [010] directions. 
Therefore, the polarization angle of the emission
lines can be used to determine the critical point $p_c$ in experiments. 
The above picture is also correct for QDs with $C_{2v}$ symmetry, 
where $\kappa = 0$, and the results
in Fig. \ref{fig-lowbounds} (a) are then
reduced to the results presented in Ref. [\onlinecite{singh10}].

The theory also provides a simple way to determine the FSS lower bound
of a QD before applying the external stress. At $p=0$, 
we have $\Delta_0 = 2\sqrt{\delta^2
+ \kappa^2}$ and polarization angle 
$\tan(\theta) = {\delta\over \kappa} \pm \sqrt{1
+({\delta \over \kappa})^2}$. It is easy to show that 
$\kappa = -\Delta_0 \cdot \sin(2\theta)/2$ and 
$\delta = \Delta_0 \cdot \cos(2\theta)/2$. 
Thus by measuring the polarization angle and FSS at $p$=0, 
we can uniquely determine the values of
2$|\kappa|$ and 2$|\delta|$ which are the FSS lower bounds for the stress along
the [110] ([1$\bar{1}$0]) and [100] directions. 
For entangled photon source applications, $|\delta|$ or $|\kappa|$ need 
to be smaller than 1 $\mu$eV. 
Therefore our results provide a useful guide in selecting QDs for entangled photon
sources, i.e., one should pick QDs for which the polarization angle is as
closely aligned as possible to
the [110] or [100] direction. 

To confirm the above analysis, we perform numerical calculations of the FSS and 
polarization angle of InAs/GaAs QDs using an empirical pseudopotential
method \cite{wang99b, williamson00}. 
We model the InAs/GaAs QDs by embedding the InAs dots into a
60$\times$60$\times$60
8-atom GaAs supercell. 
To calculate the exciton energies and their FSS, we first have to obtain
the single-particle energy levels and wavefunctions 
by solving the Schr\"{o}dinger equation,
\begin{equation}
\left[ -{1 \over 2} \nabla^2
+ V_{ps}({\bf r}) \right] \psi_i({\bf r})
=\epsilon_i \;\psi_i({\bf r}) \; ,
\label{eq:schrodinger}
\end{equation}
where $V_{\rm ps}({\bf r})=V_{\rm SO}+\Sigma_{i}\Sigma_{\alpha}
\upsilon_{\alpha}({\bf r}-{\bf R}_{i,\alpha}) + \mathcal{U}_{\text{piezo}}$ is
the total pseudopotential and
$\upsilon_{\alpha}({\bf r}-{\bf R}_{i,\alpha})$ is the local {\it screened} atomic
potential at the equilibrium atom position ${\bf R}_{i,\alpha}$ 
obtained by minimizing the total strain energies under the 
given stress \cite{singh10} using the valence 
force field method \cite{keating66}.
$V_{\rm SO}$ is the spin-orbit interaction, and 
$\mathcal{U}_{\text{piezo}}$ is the piezoelectric potential \cite{bester06b}. 
The single particle energy levels are calculated using a linear combination of bulk 
bands method \cite{wang99b}.
The exciton energies are then calculated via many-particle
configuration interaction (CI) method \cite{franceschetti99},
in which the exciton wavefunctions are 
expanded in Slater determinants constructed
from all confined electron 
and hole single-particle states.

\begin{figure}
\centering
\includegraphics[width=3.2in]{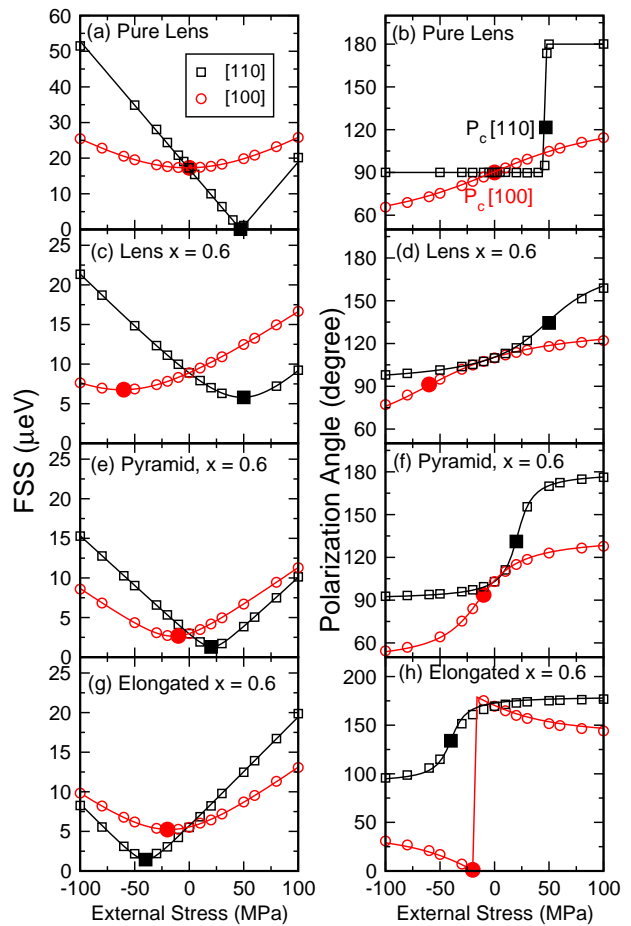}
\caption{(Color online) Left panel: The $\text{FSS}$ as a function of $p$ for
(a) pure lens-shaped, (c) alloy lens-shaped, (e) alloy pyramidal and (g)
alloy elongated InAs/GaAs QDs. Right panel: Corresponding exciton polarization angle as
a function of $p$.
The open squares (circles) are the results for stress along the [110] ([100])
direction calculated from the atomistic pseudopotential method, whereas the solid 
lines are the results of the theory. Solid squares and circles indicate critical
stresses.  
}
\label{fig-all}
\end{figure}

\begin{table}
\caption{Parameters for In$_{x}$Ga$_{1-x}$As/GaAs QDs under uniaxial stress.
${\bf n}$ is the direction of external stress, $\alpha$ and $\beta$
are in units of $\mu$eV/MPa, $\kappa$ and $\delta$ are in the units of
$\mu$eV, the critical stress $p_c$ is in
MPa and the lower bounds $\delta_{\text{b}}$ is in units of $\mu$eV.
The unit for base diameter $D$ and height $h$ of the dots is nm.}
\begin{center}
\begin{tabular}{lccccccc}\hline \hline 
QDs    & ${\bf n}$  & $\alpha$ & $\beta$  & $\delta$  & $\kappa$ & $p_c$ & $\delta_{\text{b}}$\\ \hline
Lens ($x$=0)    & [110]         & 0.36  & 0    & -8.46  & 0      & 47  & 0 \\
$D$=20, $h$=3.5 	     & [1$\bar{1}$0] & -0.37 & 0    & -8.55  & 0      &-46  & 0 \\
                & [100]         & 0     &-0.095& -8.65  & 0      & 0   & 17.3\\\hline
Lens ($x=0.6$)   & [110]         & 0.14  & 0    & -3.36  & -2.90  & 49  & 5.8\\
$D$=25, $h$=3.5           & [1$\bar{1}$0] & -0.14 & 0    & -3.36  & -2.86  &-47  & 5.7 \\
	          & [100]         & 0     &-0.047&-3.38   &-2.90   &-62  & 6.8\\
Pyramid ($x$=0.6) & [110]         & 0.13  & 0    & -1.32  & -0.64  & 21  & 1.3\\
$D$=25, $h$=3.5    & [1$\bar{1}$0] & -0.13 & 0    & -1.41  & -0.61  &-22  & 1.2\\
            & [100]         & 0     &-0.048&-1.32   &-0.69   & 14  & 2.6\\
Elongated ($x$=0.6)  & [110]         & 0.14  & 0    & 2.76   & -0.72  &-40  & 1.4\\
$D_{[1\bar{1}0]}$=26,    & [1$\bar{1}$0] & -0.14 & 0    & 2.73   & -0.78  & 38  & 1.6\\
$D_{[110]}$=20, $h$=3.5 	         & [100]         & 0     &-0.051& 2.62   & -0.92  &-18  & 5.2\\
\hline \hline
\end{tabular}
\end{center}
\label{table-all}
\end{table}

The change in FSS under external stress is purely an atomistic effect, because
the macroscopic shapes of the dots change little (less than 0.1\%) 
under such stresses and should not affect the FSS.
Furthermore, we find that including
piezoelectricity or not gives essentially the same results, 
suggesting that piezoelectricity is not
responsible for the FSS change. In fact, the change in FSS is due to changes in
the underlying atomic structure.  
We have calculated the FSS of more than 13 dots under external stresses along
the [110], [1$\bar{1}$0] and [100] directions. The behaviors of FSS are almost symmetric
for stresses along the [110] and [1$\bar{1}$0] directions, i.e., the effects of tensile stress along
the [110] direction is almost identical to the effects of compression along the
[1$\bar{1}$0] direction. 
The results for some typical dots are shown in Fig. \ref{fig-all}, whereas 
the geometry and other parameters of these dots are listed in Table.~\ref{table-all}.
In the left panels of Fig. \ref{fig-all}, we plot the FSS
vs $p$ along the [110] (black square) and [100] (red circle) directions. 
The solid lines are fitted from theory using Eq. (\ref{eq-FSS}).
The right panels show the corresponding polarization angle 
$\theta$ vs $p$ where the solid lines are the theoretical 
predictions using Eq. (\ref{eq-theta}). 
We use $\beta$=0 ($\alpha$=0) for $p$ along [110] ([100]) direction. As can be seen,
the agreement between numerical calculations and theory is remarkable. 

(1) {\it Pure InAs/GaAs QDs with $C_{2v}$ symmetry.} In Fig. \ref{fig-all} (a) we show the FSS vs $p$ 
along different directions for a pure lens-shaped QDs with base $D$ =20 nm and height $h$=3.5
nm. When the stresses are directed along the 
[110] and $[1\bar{1}0]$ directions, the
FSS can be tuned exactly to zero \cite{singh10}. 
The polarization angle $\theta$ is constant (90$^\circ$) below $p_c$ = 47 MPa and jump to 180$^\circ$
after $p_c$ as shown in Fig. \ref{fig-all} (b). 
However, if the stress is along the $[100]$ direction, the FSS can not be tuned
to zero, in agreement with previous results \cite{singh10}. The
polarization angle rotates following Eq. (\ref{eq-theta}) as seen
in Fig. \ref{fig-all} (b). At $p=p_c$, the polarization angle is 90$^\circ$.
We also calculate pure pyramidal and elongated QDs, 
which do not have macroscopic cylindrical symmetry but still retain
$C_{2v}$ symmetry, and find similar features.

(2) {\it Alloy In$_{0.6}$Ga$_{0.4}$As/GaAs QDs with $C_1$ symmetry.} 
For alloy dots, the symmetry is lowered to $C_1$. The FSS has lower bound
under the uniaxial stress \cite{singh10}, as shown in Fig. \ref{fig-all} (c) 
(e) and (g), for different dot geometries and sizes. 
The corresponding parameters are summarized in Table \ref{table-all}. 
The stress dependence of the polarization angles is 
also in excellent agreement with theory as shown in
Fig. \ref{fig-all} (d), (f) and (h) for the three dots. 
All three dots have polarization angle $\theta_c$=135$^\circ$ at $p_c$ if
the stress is along the [110] direction, and  $\theta_c$=90$^\circ$ (or
0$^\circ$) if the stress is along the [100] direction, as predicted by the theory.    
Among the three alloy dots, the lens-shaped QD [Fig. \ref{fig-all} (c)] has
the largest lower bounds $\sim$ 7 $\mu$eV at $p_c\sim$ 62 MPa along the [100]
direction. At $p$=0, the lens-shaped QD has $\theta$ = 110$^\circ$, compared
to $\theta$ = 103$^\circ$ for the pyramidal dot and 169$^\circ$ for the
elongated dot. The polarization angle of 
the lens-shaped QD deviates from the [110] (or [1$\bar{1}$0]) direction most, and hence
has the largest FSS at $p_c$, as predicted by the theory. 
We obtain similar results when the stress in along
the [100] direction.

In the calculations, we find that $\alpha$ is not very sensitive to the QD
shape, but changes with alloy compositions. For example, for pure dots,
$|\alpha| \sim$ 0.2 - 0.4 $\mu$eV/MPa for $p$ along the [110] direction, 
whereas for In composition $x$=0.6, $|\alpha|$ reduces to 0.1 - 0.2 $\mu$eV/MPa. 
$\beta$ also 
has similar features (for $p$ along the [100] direction), 
with $|\beta| \sim$ 0.05 - 0.1 $\mu$eV/MPa for 
pure dots, and  $|\beta| \sim$ 0.04 - 0.05 $\mu$eV/MPa for alloy dots with $x$=0.6.
We also calculate alloy dots of the same geometry but with 
different alloy distributions \cite{mlinar09} and find that the
alloy distribution does not significantly change the values of $\alpha$ and $\beta$. 
In contrast, $\delta$, $\kappa$, and the polarization angle $\theta$ at $p$=0,
change dramatically from dot to dot, in agreement with recent experiments \cite{favero05}. 
However, in all cases, the behaviors of the FSS and polarization angle under stress
are in excellent agreement with our theoretical predictions.

To conclude, we have established a general relationship between the
asymmetry in QDs, the exciton polarization angle, 
and the FSS under uniaxial stress. 
We showed that the FSS lower bound under external stress 
can be predicted by the polarization angle and FSS under zero stress.
The critical stress can also be determined by monitoring the 
change in exciton polarization angle.
The work therefore provides a useful guide in selecting 
QDs with smallest FSS which is crucial for entangled photon sources 
applications.

The authors thank A. J. Bennett for bringing
Ref. \onlinecite{bennett10} to our attention.
LH acknowledges the support from the Chinese National
Fundamental Research Program 2011CB921200 and
National Natural Science Funds for Distinguished Young Scholars.

{\it Note added:} After submitted the paper, 
we became aware of Ref. \onlinecite{bennett10}. 
There, the exciton FSS of
InAs/GaAs QDs is tuned via an electric field along the 
[001] direction, which has the
same symmetry as applying stress along the [110] and [1-10]
directions (i.e., the case of $\beta$=0). 
Determining the degree of agreement between the present
theory and the experiment of Ref. \onlinecite{bennett10} 
is a promising avenue for future research. 


\begin{thebibliography}{21}
\bibitem[{foo()}]{footnote1}
\bibinfo{note}{The linear polarization is protected by the
  time-reversal symmetry. If the time-reversal symmetry is broken, the
  parameters are not always real and the polarization will become circular.}
\end{thebibliography}

\end{document}